\documentclass[aps,preprint,prd,showpacs,nofootinbib]{revtex4}

\usepackage{amsmath,amssymb}
\usepackage{graphicx,subfigure}
\usepackage{color,multirow}
\usepackage[colorlinks,linkcolor=magenta,anchorcolor=cyan,citecolor=blue,plainpages=false]{hyperref}

\hypersetup{colorlinks=true,
    breaklinks=true,
    pdfstartview=Fit,
    linkcolor=blue,
    citecolor=blue,
    urlcolor=blue}

\def\be{\begin{equation}}
    \def\ee{\end{equation}}
\def\ba{\begin{eqnarray}}
    \def\ea{\end{eqnarray}}

\begin{document}

\title{Imprint of swampland-inspired coupled early dark energy}

\author{Hao Wang$^{1,2} $\footnote{\href{wanghao187@mails.ucas.ac.cn}{wanghao187@mails.ucas.ac.cn}}}
\author{Yun-Song Piao$^{1,2,3,4} $ \footnote{\href{yspiao@ucas.ac.cn}{yspiao@ucas.ac.cn}}}

    \affiliation{$^1$ School of Fundamental Physics and Mathematical
        Sciences, Hangzhou Institute for Advanced Study, UCAS, Hangzhou
        310024, China}

    \affiliation{$^2$ School of Physics Sciences, University of
        Chinese Academy of Sciences, Beijing 100049, China}

    \affiliation{$^3$ International Center for Theoretical Physics
        Asia-Pacific, Beijing/Hangzhou, China}

    \affiliation{$^4$ Institute of Theoretical Physics, Chinese
        Academy of Sciences, P.O. Box 2735, Beijing 100190, China}

    \begin{abstract}
Inspired by the Swampland Distance Conjecture, we investigate the
cosmological implications of a fractional coupling between dark
matter (DM) and early dark energy (EDE) in light of the recent
DESI DR2 BAO data. We use a conditional normalizing flow network
to efficiently sample the high-dimensional parameter space, and
perform a joint analysis of Planck CMB data, DESI DR2 BAO,
PantheonPlus supernovae and SH0ES. We find that the detailed
construction of the EDE potential beyond the mere existence of an
EDE component possibly alter cosmological constraints on late-time
dark energy when the coupling between DM and EDE is considered.
    \end{abstract}

    \maketitle
    \tableofcontents

\section{Introduction}
The recent release of the Dark Energy Spectroscopic Instrument
(DESI) Data Release 2 (DR2) has provided unprecedented
measurements of baryon acoustic oscillations (BAO) across a wide
redshift range
$0.1<z<4.2$\cite{DESI:2024lzq,DESI:2024mwx,DESI:2024uvr,DESI:2025zgx},
offering new insights into the expansion history of the Universe.
Combined with cosmic microwave background (CMB) data from Planck
and supernova observations, these data hint at a possible
$\gtrsim3\sigma$ preference for evolving dark energy (DE) over the
cosmological constant (CC), particularly in the context of
$w_0w_a$CDM with CPL (Chevallier–Polarski–Linder)
parameterization\cite{Chevallier:2000qy,Linder:2002et}. These
results are under active investigation,
e.g.\cite{Luongo:2024fww,Cortes:2024lgw,Carloni:2024zpl,Colgain:2024xqj,Giare:2024smz,Wang:2024dka,Yang:2024kdo,Park:2024jns,Shlivko:2024llw,Dinda:2024kjf,Seto:2024cgo,Bhattacharya:2024hep,Roy:2024kni,Wang:2024hwd,Notari:2024rti,Heckman:2024apk,Gialamas:2024lyw,Orchard:2024bve,Colgain:2024ksa,Wang:2024sgo,Li:2024qso,Ye:2024ywg,Giare:2024gpk,Dinda:2024ktd,Jiang:2024viw,Alfano:2024jqn,Jiang:2024xnu,Sharma:2024mtq,Ghosh:2024kyd,Reboucas:2024smm,Pang:2024qyh,Wolf:2024eph,RoyChoudhury:2024wri,Arjona:2024dsr,Wolf:2024stt,Giare:2024ocw,Wang:2024tjd,Alestas:2024eic,Carloni:2024rrk,Bhattacharya:2024kxp,Specogna:2024euz,Li:2024qus,Ye:2024zpk,Pang:2024wul,Akthar:2024tua,Colgain:2024mtg,daCosta:2024grm,Chan-GyungPark:2025cri,Sabogal:2025mkp,Du:2025iow,Ferrari:2025egk,Jiang:2025ylr,Peng:2025nez,Jiang:2025hco,Feng:2025mlo,Hossain:2025grx,Chakraborty:2025syu,Borghetto:2025jrk,Pan:2025psn,Pang:2025lvh,Wang:2025ljj,Kessler:2025kju,Yang:2025mws,Wolf:2025jed,RoyChoudhury:2025dhe,Specogna:2025guo,Ye:2025ark,Cheng:2025lod,Ling:2025lmw,Wolf:2023uno,Wang:2025dtk,Li:2025eqh,Cline:2025sbt,Gialamas:2025pwv,Li:2025dwz,Lee:2025pzo,Ishak:2025cay,Wang:2025vtw,Wang:2025znm,Zhou:2025nkb,RoyChoudhury:2025iis,Pedrotti:2025ccw,Wang:2026kbg,Wolf:2023uno}.

The nature of dark energy and its possible connection to other
dark sectors remains one of the most profound mysteries in modern
cosmology. The Hubble tension\cite{Verde:2019ivm,Riess:2019qba},
which reveals the stark contrast between the Hubble constant
measurements of $H_0 \sim 73$ km/s/Mpc from the SH0ES
collaboration, which uses Cepheid-calibrated Type Ia supernovae
\cite{Breuval:2024lsv}, and $H_0 \sim 67$ km/s/Mpc inferred by the
Planck collaboration based on the standard $\Lambda$CDM model
using their cosmic microwave background (CMB) data
\cite{Planck:2018vyg}, has motivated many solution
attempts\cite{Knox:2019rjx,Perivolaropoulos:2021jda,DiValentino:2021izs,Vagnozzi:2023nrq},
among which Early Dark Energy (EDE)
\cite{Karwal:2016vyq,Poulin:2018cxd,Smith:2019ihp} is a leading
pre-recombination scenario. By becoming non-negligible just only
before recombination, EDE suppressed the sound horizon and raises
$H_0$, with an anti-de Sitter phase around recombination further
enhancing its effect (AdS-EDE) \cite{Ye:2020btb,Ye:2020oix}. See
\cite{Agrawal:2019lmo,Lin:2019qug,Niedermann:2019olb,Berghaus:2019cls,Sakstein:2019fmf,Zumalacarregui:2020cjh,Ballesteros:2020sik,Braglia:2020iik,Ballardini:2020iws,Gogoi:2020qif,Braglia:2020bym,Gonzalez:2020fdy,Braglia:2020auw,CarrilloGonzalez:2020oac,Adi:2020qqf,Ye:2021nej,Karwal:2021vpk,Jiang:2021bab,Gomez-Valent:2021cbe,Poulin:2021bjr,Niedermann:2021vgd,Wang:2022jpo,Smith:2022hwi,Jiang:2022uyg,Gomez-Valent:2022bku,Escudero:2022rbq,Herold:2022iib,Goldstein:2023gnw,Eskilt:2023nxm,Odintsov:2023cli,Sharma:2023kzr,Fu:2023tfo,Pedreira:2023qqt,Khalife:2023qbu,Forconi:2023hsj,Wang:2024jug,Giare:2024akf,Chatrchyan:2024xjj,Jiang:2024nha,Forconi:2025cwp,Piras:2025eip,Stahl:2025czl,Jiang:2025hco,Forconi:2025zzu,Smith:2025grk,Yashiki:2025loj,Toda:2025kcq,Yin:2026gss,Gonzalez-Fuentes:2026rgu,Jhaveri:2026bla}
for other promising scenarios of EDE models. This
pre-recombination resolution of Hubble tension suppress the
preference of DESI for the evolving DE, which suggests that the
evidence against CC is likely to disappear when certain resolution
of Hubble tension is
adopted\cite{Wang:2024dka,Wang:2025djw,Pang:2024wul,Pang:2025lvh}.

Parallel to observational developments, theoretical considerations
from string theory, specifically the Swampland Distance Conjecture
(SDC)\cite{Ooguri:2006in}, suggest that scalar fields traversing
large field distances can induce exponential couplings to other
sectors, such as dark matter (DM). This motivates the
consideration of coupled EDE models, where a scalar field
responsible for resolving the Hubble tension also couples to DM,
leading to a time-dependent DM
mass\cite{Agrawal:2019dlm,McDonough:2021pdg,Wang:2022bmk}.

In this work, we investigate the impact of such swampland-inspired
coupled EDE on the evolving dark energy preferred by DESI BAO
data. According to \cite{McDonough:2021pdg,Wang:2022bmk}, we
consider two distinct EDE realizations: the axion-like EDE
model\cite{Poulin:2018cxd}, where the scalar field oscillates and
settles at the bottom of its potential, and the AdS-EDE
model\cite{Ye:2020btb}, where the scalar field continues to evolve
at late time along a runaway AdS potential. To efficiently explore
the high-dimensional parameter space, we employ a conditional
normalizing flow
network\cite{Dinh:2014mzt,Rezende:2015ocs,Kobyzev:2019ydm,Papamakarios:2019fms}
to model posterior distributions, avoiding the computational
bottlenecks and potential volume effects of MCMC methods. Compare
constraints obtained with DESI DR2 BAO data against previous
results using SDSS BAO\cite{Wang:2022bmk}, we find that the
detailed construction of the EDE potential beyond the mere
existence of an EDE component possibly alter cosmological
constraints on late-time dark energy when the coupling between DM
and EDE is considered.

\section{Method and Datasets}

\begin{figure}[htbp]
        {\includegraphics[width=0.8\textwidth]{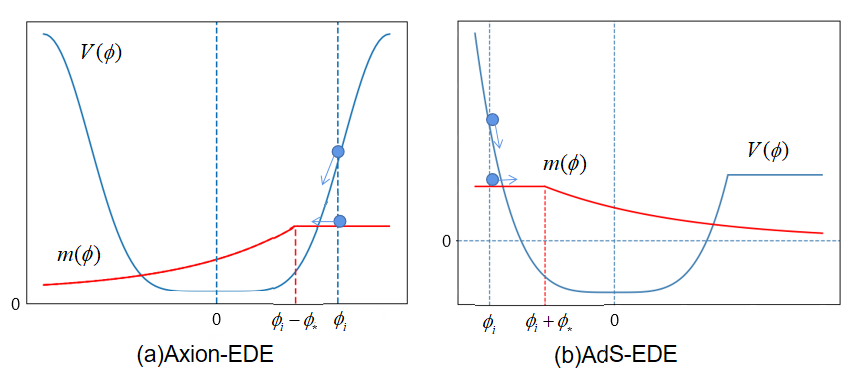}}
\caption{\label{fig1}A sketch of the EDE potential $V(\phi)$ and
$m(\phi)$ in axion-like EDE and AdS EDE models, respectively.
Initially the field sits at $\phi_i$, after its excursion
$|\Delta\phi|>\phi_*$, the mass of DM will be exponentially
lightened with the evolution of $\phi$. }
    \end{figure}

Following \cite{McDonough:2021pdg,Wang:2022bmk}, we consider the
couple of DM with EDE inspired by SDC\cite{Ooguri:2006in}. The
dark matter is modelled as a population of non-relativistic Dirac
fermions $\psi$, \be {\cal L}_{int}\sim
 -m_{cdm}(\phi)\bar{\psi} \psi,\label{L}\ee \be
 m_{cdm}(\phi)=f_*m(\phi)+(1-f_*)m_i,\quad with \quad 0\leqslant f_*\leqslant 1, \label{mdm}\ee
 \be
m(\phi)=m_i e^{-c({|\Delta\phi|-\phi_*\over M_{pl}})},\quad for
\quad |\Delta\phi|\geq\phi_*.
    \label{mphi}\ee
where $m_i=\mathrm{const.}$ is the initial mass of DM,
$|\Delta{\phi}|=|\phi-\phi_i|$ (see Fig.\ref{fig1}), $\phi_*$
signals the insensitivity of DM on a shift of $\phi$ within
$|\Delta{\phi}|<\phi_*$, and $c$ is the coupling intensity. Here,
when $c=0$, we have $m_{cdm}=m_i$ (the standard EDE+$\Lambda$CDM
model is recovered).

In non-relativistic limit, we have $\rho_{cdm}= nm_{cdm}(\phi)$,
so \be \rho_{cdm}= nf_*m(\phi)+n(1-f_*)m_i,\ee with $n$ being the
number density, which suggests that $f_*$ is actually equivalent
to the fraction of DM coupled to EDE. The evolution of EDE
is rewitten as $\phi^{\prime\prime}+2{\cal
H}\phi^\prime+a^2V_{\phi}=-a^2\frac{d\rho_{cdm}}{d\phi}$, while
the continuity equation for DM is
    \be
    \rho_{cdm}^\prime+3{\cal H}\rho_{cdm}=\phi^\prime\frac{d\rho_{cdm}}{d\phi},
    \label{rhocdm}\ee
where the \textit{prime} is the derivative with respect to
$d\eta=dt/a$, and ${\cal H}={a^\prime/a}$. Integrating
Eq.(\ref{rhocdm}), we have
    \ba
\rho_{cdm}(a)=
\frac{3M_{pl}^2H_0^2\Omega_{cdm}}{a^3}[1-f(\phi_0)+\frac{m(\phi)}{m(\phi_0)}f(\phi_0)],
\label{rhocdm1}\ea where
$f(\phi)=\frac{{m(\phi)}f_*}{{m(\phi)}f_*+m_i(1-f_*)}$, and
$\phi_0$ is the present-day value of $\phi$.

In our analysis, we consider the $w_0w_a$CDM model (well-known CPL
parameterisation for DE) where the state equation of DE is
\cite{Chevallier:2000qy,Linder:2002et} \be
w_\mathrm{DE}=w_0+w_a{z\over 1+z}.\ee

We consider two EDE models, axion-like\cite{Poulin:2018cxd} and
AdS EDE\cite{Ye:2020btb} respectively\footnote{The corresponding
    cosmological codes are available at: axion-like EDE
    (\url{https://github.com/PoulinV/AxiCLASS}) and AdS-EDE
    (\url{https://github.com/genye00/class_multiscf}).}. In axion-like EDE
model, an axion field with $V(\phi)\propto(1-\cos[\phi/f])^3$ is
responsible for EDE \cite{McDonough:2022pku,Kojima:2022fgo}, which
starts to oscillate at the redshift $z_c\sim 3000$, see
Fig.\ref{fig1}(a), $\phi-\phi_i<0$ for $\phi_i>0$, so
$d\rho_{cdm}/d\phi=c\rho_{cdm}f(\phi)/{M_{pl}}$. The scalar field
will eventually settle at the bottom of its potential, so
$\phi_0=0$. In the synchronous gauge, with $\rho_{cdm}$ in
Eq.(\ref{rhocdm1}), the perturbations equations have been derived
fully in Ref.\cite{McDonough:2021pdg}. The parameter space is
$\{\omega_b,\omega_{cdm},H_0,\ln({10^{10}}A_s),n_s,\tau_{reio},{\log_{10}}{a_c},f_{ede},\phi_i,w_0,w_a,c,\phi_*,f_*\}$.
However, in AdS-EDE model we have
$V(\phi)=V_0(\phi/M_p)^{4}-V_{ads}$, which is glued to a
cosmological constant $V(\phi)=\Lambda$ by interpolation
($V_{ads}>0$ is the AdS depth), and $\phi-\phi_i>0$, see
Fig.\ref{fig1}(b), so
$d\rho_{cdm}/d\phi=-c\rho_{cdm}f(\phi)/{M_{pl}}$. This suggests
that $\phi_0$ must be obtained by solving the equation of motion,
so it is not convenient to use Eq.(\ref{rhocdm1}). Integrating
Eq.(\ref{rhocdm}), instead we have \ba \rho_{cdm}^{(AdS)}(a)&=&
\frac{3M_{pl}^2H_0^2{\Omega}_{cdm}}{a^3}[1-f(\phi_0)+(1-f(\phi_0)){f_*\over
    1-f_*} \frac{m(\phi)}{m(\phi_{i})}]\nonumber\\
&=&
\frac{3M_{pl}^2H_0^2{\tilde\Omega}_{cdm}}{a^3}[1+f_*^{(AdS)}\frac{m(\phi)}{m(\phi_{i})}],
\label{rhocdm3}\ea where ${\tilde
\Omega}_{cdm}=\Omega_{cdm}(1-f(\phi_{0}))$ is defined to absorb
$\phi_0$ and $f_*^{(AdS)}= f_*/(1-f_*)$. The parameter space is
$\{\omega_b,\omega_{cdm},H_0,\ln({10^{10}}A_s),n_s,\tau_{reio},\ln(1+z_c),
f_{ede},w_0,w_a,c,\phi_*,f_*^{(AdS)}\}$. In order to have a
significant AdS phase while make the field able to climb out of
the AdS well, we fixed $V_{ads}$ by setting $ V_{ads}=0.26\times
10^{4} \left(\rho_\text{m}(z_c)+\rho_\text{r}(z_c)\right)$, as in
Ref.\cite{Ye:2020btb}.

Here, we use \textbf{DESI DR2} BAO data, which consists of bright
galaxies, LRGs, ELGs, quasars and Ly$\alpha$ Forest samples at the
redshift region $0.1<z<4.2$ \cite{DESI:2025zgx}, which is
consistent with SDSS and DESI DR1 \cite{DESI:2024mwx}. To compare
with the previous results using SDSS BAO data\cite{Wang:2022bmk},
we use the \textbf{Planck PR3} dataset (low-T \texttt{Commander},
low-E \texttt{SimALL} and Planck 2018 high-$l$ TT, TE, EE spectra,
and reconstructed CMB lensing spectrum
\cite{Planck:2018vyg,Planck:2019nip,Planck:2018lbu}),
\textbf{PantheonPlus} (consisting of 1701 light curves of 1550
spectroscopically confirmed Type Ia SN coming from 18 different
surveys \cite{Scolnic:2021amr}) and \textbf{SH0ES}
Cepheid-calibrated SN1a magnitude (equivalently SH0ES prior
\cite{Riess:2021jrx}).

\begin{figure}[htbp]
    {\includegraphics[width=\textwidth]{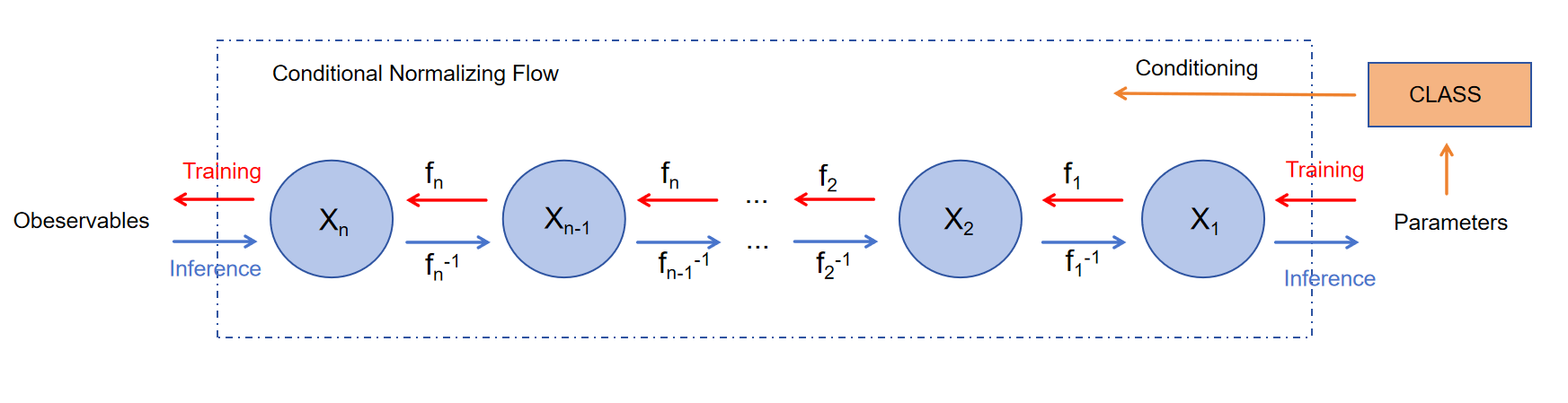}}
\caption{\label{CNF}A sketch of our conditional normalizing flow
network. Red and blue solid arrows represent the directions of
training and inference respectively. $f_n$ denotes the
transformation after each training step which maps from $X_{n-1}$
to $X_n$.}
\end{figure}

In this work, to fully explore the huge parameter space consisting
of 13/14 freedom degrees, we model the posterior distributions
using a conditional normalizing flow
network\cite{Dinh:2014mzt,Rezende:2015ocs,Kobyzev:2019ydm,Papamakarios:2019fms}.
In addition to the significant computational speed-up, our
approach avoids the explicit construction of complex likelihood
functions and potential volume
effect\cite{Gomez-Valent:2022hkb,Schoneberg:2021qvd,Herold:2021ksg}
in MCMC analysis. We utilize the implementation of this structure
provided by \texttt{normflows} (\url{
https://github.com/VincentStimper/normalizing-flows})\cite{Stimper2023}.
The general structure of our pipeline is depicted in Fig.\ref{CNF}
which displays the input/output relations. Normalizing flows
facilitate efficient sampling and density estimation by expressing
the distribution as a sequence of mappings, or ``flows”,
$f:u\rightarrow \Lambda$, which maps from a standard normal
distribution $u$ to the parameter space of EDE models. In our
training process, the flow mapping is made dependent on data and
extra observables produced by EDE models. The conditional
probability distribution
$q(\Lambda|d)=\mathcal{N}(0,1)f^{-1}(\Lambda|d)|\det
f^{-1}(\Lambda|d)|$ where $d$ denotes the train set. Once trained,
drawing posterior samples for a new observation requires only a
single forward pass through the flow, yielding thousands of
independent samples in accelerating parameter inference for EDE
cosmologies.

\section{Results}

As shown in Table \ref{tab1}, for the axion-like EDE model with
the \texttt{PlanckPR3+DESI+Pantheon+SH0ES} dataset, we find
$c\sim0$ within the $1\sigma$ region. This result is consistent
with previous findings using pre-DESI BAO data
\cite{McDonough:2021pdg,Wang:2022bmk}, indicating that the
coupling defined in Eq. (\ref{L}) is disfavored. Similar to Ref.
\cite{Wang:2022bmk}, the inclusion of this coupling does not
significantly alleviate the $S_8$ tension, even when a CPL-like
dark energy component is considered.

\begin{figure}[htbp]
    {\includegraphics[width=0.8\textwidth]{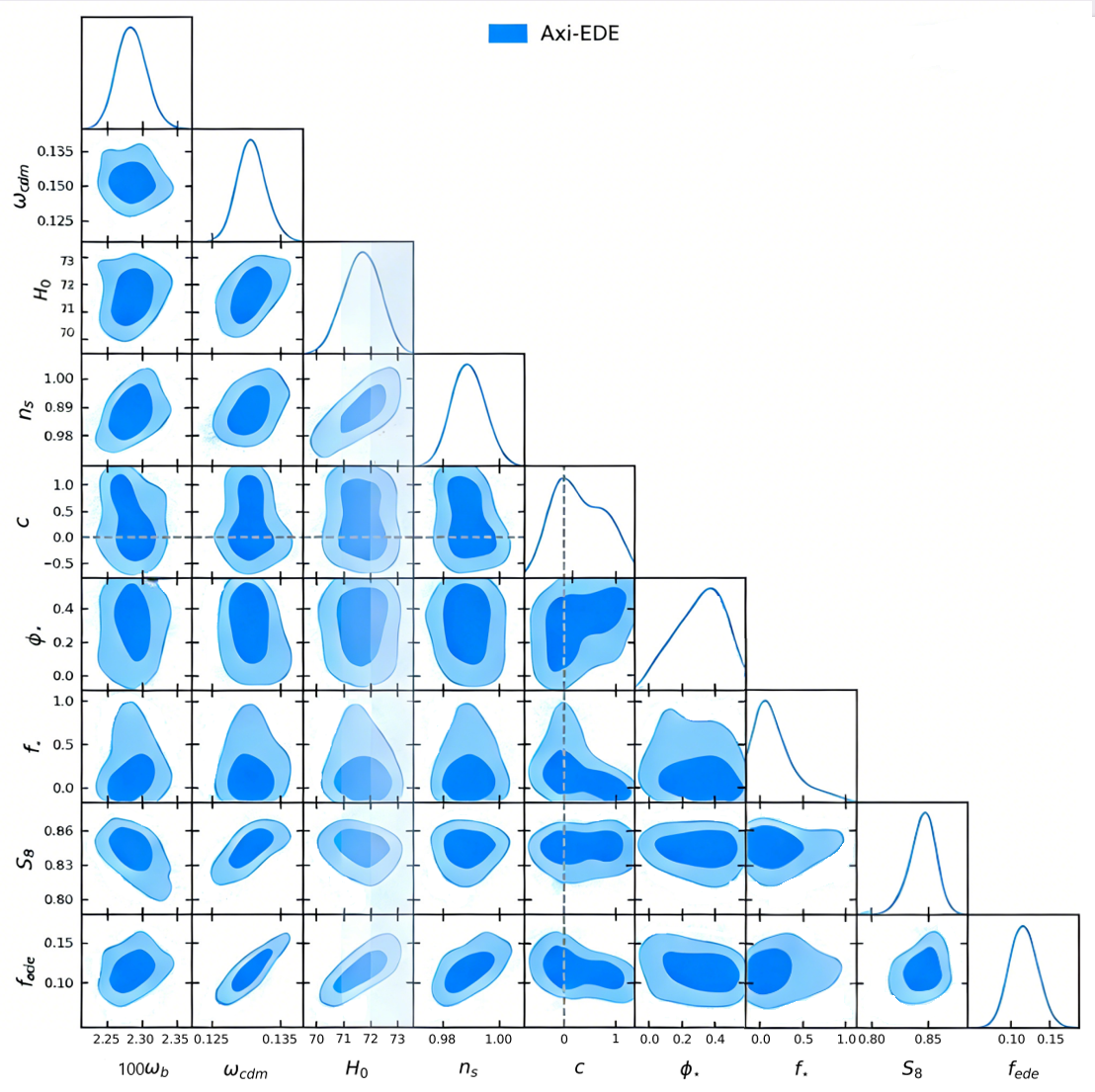}}
\caption{\label{fig2} Posterior distributions for axion-like EDE
with coupling (\ref{mphi}) in fit to \texttt{PlanckPR3+DESI+Pantheon+SH0ES} datasets. The shadows correspond to the 1$\sigma$ and 2$\sigma$ regions of $H_0$ in light of
SH0ES \cite{Riess:2021jrx}.}
    \end{figure}

\begin{figure}[htbp]
    {\includegraphics[width=0.8\textwidth]{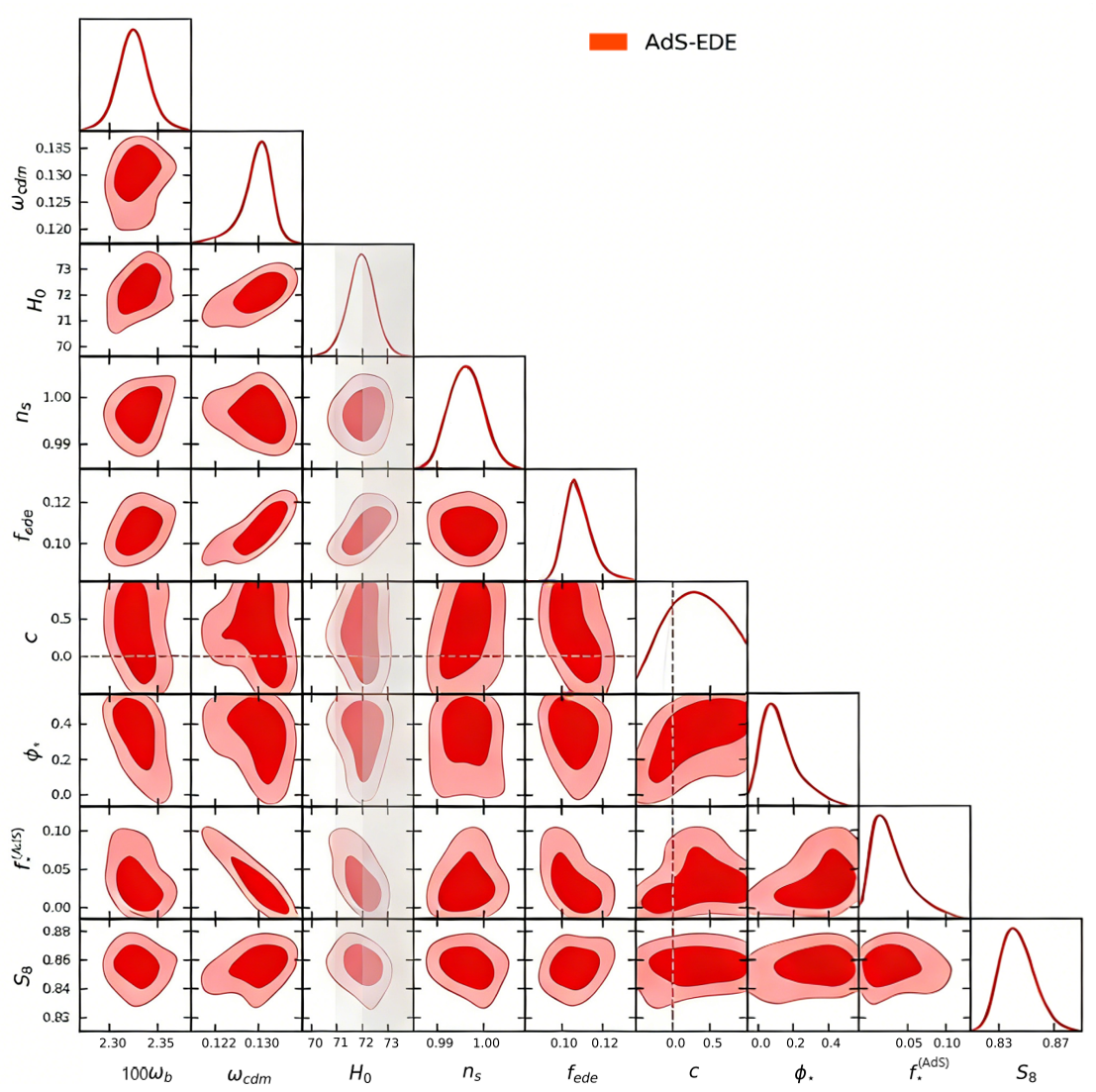}}
    \caption{\label{fig3}Posterior distributions for AdS-EDE with
        coupling (\ref{mphi}) in fit to \texttt{PlanckPR3+DESI+Pantheon+SH0ES} datasets. The shadows correspond to the 1$\sigma$ and 2$\sigma$ regions of $H_0$ in light of
        SH0ES \cite{Riess:2021jrx}.}
\end{figure}

The posterior distributions of the parameters are presented in
Figs. \ref{fig2} and \ref{fig3}. In both EDE models, a larger
$\phi_*$ is preferred, suggesting that an evolving dark energy
component favors an earlier onset of coupling. While the best-fit
value of $c$ is negative, inconsistent with SDC in the axion-like
EDE scenario, the possibility of $c > 0$ remains not ruled out.

In contrast, the AdS-EDE model yields a larger mean value of $c$
compared to the axion-like EDE case, with $f_* > 0$ preferred at
the $1\sigma$ level. Nevertheless, $c = 0$ remains consistent
within $1\sigma$. This behavior can be attributed to the runaway
potential of the AdS-EDE model, where the field undergoes a
nonzero excursion at late times, allowing the coupling between EDE
and dark matter to be constrained by low-redshift observations.

    \begin{table}[htbp]
        \scalebox{0.85}{
        \begin{tabular}{|c|c|c|c|c|}
            \hline
            \multirow{2}{*}{Parameters}&\multicolumn{2}{|c|}{Axion-EDE}&\multicolumn{2}{|c|}{AdS-EDE}\\\cline{2-5}
            &SDSS&DESI&SDSS&DESI\\
            \hline
            100$\omega_b$&$2.284(2.286)\pm0.020$&$2.289(2.293)\pm0.024$&$2.328(2.327)\pm0.014$&$2.334(2.339)\pm0.021$\\
            $\omega_{cdm}$&$0.130(0.129)\pm0.002$&$0.130(0.131)\pm0.003$&$0.130(0.130)\pm0.004$&$0.132(0.132)\pm0.002$\\
            $H_0$&$71.66(70.87)\pm0.63$&$71.73(72.11)\pm0.81$&$72.01(72.05)\pm0.51$&$72.21(72.53)\pm0.71$\\
            ln($10^{10}$$A_s$)&$3.060(3.051)\pm0.013$&$3.053(3.049)\pm0.015$&$3.076(3.088)\pm0.014$&$3.071(3.078)\pm0.015$\\
            $n_s$&$0.989(0.983)\pm0.006$&$0.987(0.989)\pm0.008$&$0.996(0.997)\pm0.004$&$0.997(0.999)\pm0.008$\\
            $\tau_{reio}$&$0.058(0.057)\pm0.006$&$0.054(0.053)\pm0.008$&$0.058(0.060)\pm0.008$&$0.054(0.054)\pm0.008$\\
            $f_{ede}$&$0.116(0.101)\pm0.017$&$0.111(0.104)\pm0.018$&$0.106(0.100)\pm0.008$&$0.117(0.108)\pm0.010$\\
            \hline
            $w_0$&-&$-0.846(-0.835)\pm0.059$&-&$-0.869(-0.862)\pm0.063$\\
            $w_a$&-&$-0.649(-0.647)\pm0.260$&-&$-0.557(-0.567)\pm0.240$\\
            \hline
            $c$&$0.289(-0.129)\pm0.472$&$0.134(-0.101)\pm0.423$&$0.367(0.302)\pm0.434$&$0.448(0.403)\pm0.513$\\
            $\phi_*$&$0.305(0.361)\pm0.147$&$0.207(0.204)\pm0.150$&$0.333(0.323)\pm0.136$&$0.234(0.278)\pm0.138$\\
            $f_*$&$0.183(0.222)\pm0.229$&$0.190(0.221)^{+0.256}_{-0.229}$&$0.030(0.008)\pm0.025$&$0.034(0.021)_{-0.025}^{+0.039}$\\
            \hline
            $S_8$&$0.845(0.844)\pm0.011$&$0.845(0.843)\pm0.012$&$0.855(0.861)\pm0.010$&$0.858(0.856)\pm0.012$\\
            \hline
        \end{tabular}}
\caption{\label{tab1}Mean(best-fit) values of axion-like EDE and AdS EDE with coupling (\ref{mphi}) in fit to \texttt{PlanckPR3+SDSS+Pantheon+SH0ES}\cite{Wang:2022bmk} and \texttt{PlanckPR3+DESI+Pantheon+SH0ES} datasets, respectively.}
    \end{table}

\begin{figure}[htbp]
    {\includegraphics[width=1\textwidth]{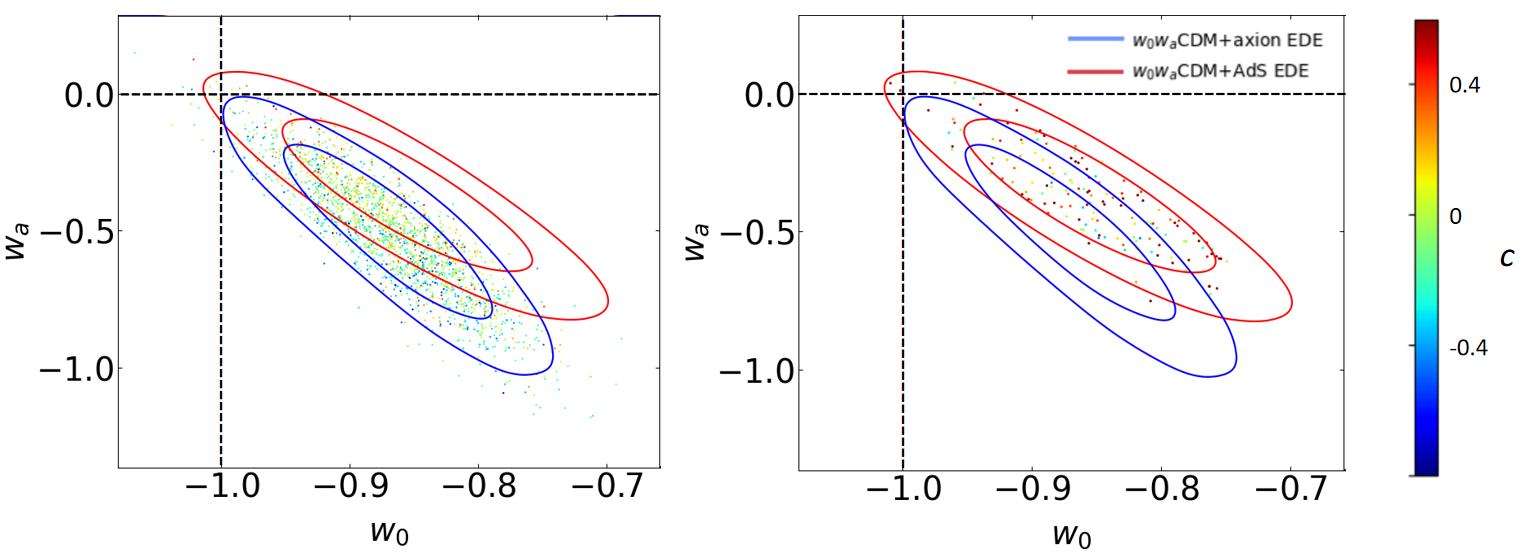}}
\caption{\label{w0wa} Posteriors for the $w_0-w_a$ scattered by
$c$ for axion-like EDE and AdS-EDE respectively, in fit to
\texttt{PlanckPR3+DESI+Pantheon+SH0ES} datasets.}
\end{figure}

To illustrate the impact of DESI data on the coupling, we present
in Fig. \ref{w0wa} the posterior distributions of the
$w_0$–$w_a$ plane, scattered by the coupling parameter $c$.
Previous studies have shown that the pre-recombination resolution
of the Hubble tension—such as that provided by EDE—suppresses
the preference for evolving dark energy implied by DESI data,
reducing the significance to $\gtrsim 3\sigma$ and rendering the
cosmological constant consistent at $\lesssim 2\sigma$
\cite{Wang:2024dka,Pang:2024wul,Pang:2025lvh}.

In this work, we find that the coupling between EDE and dark
matter favors a lower Hubble constant $H_0$, which slightly
enhances the preference for evolving dark energy to $\gtrsim
2\sigma$ in the axion-like EDE model. However, unlike the
axion-like EDE scenario where the scalar field settles at $\phi =
0$ after oscillations, the AdS-EDE model features a sustained
field excursion, with dark matter decaying after the EDE phase.

In the $w_0 w_a$CDM model, the equation-of-state parameter $w(z)$
crossing the phantom divide $w = -1$ is favored by DESI BAO data.
The regime $w(z) < -1$ corresponds to an increase in dark energy
density, which counteracts the effects of the EDE–DM coupling
and favors $c > 0$. Consequently, a preference for evolving dark
energy at the $\lesssim 2\sigma$ level persists in the AdS-EDE
model with a coupled dark matter component.

    \section{Conclusions}

In this work, motivated by the SDC, we investigate the impact of a
fractional coupling between DM and EDE on the evolving dark energy
preferred by recent DESI BAO data. Following the approach of Ref.
\cite{Wang:2022bmk}, using a conditional normalizing flow network
to model the posterior distributions, we find that the inclusion
of such a coupling does not significantly alleviate the $S_8$
tension, even when a CPL-like DE component is considered.

Comparing the constraints on the $w_0$–$w_a$ plane obtained from
combined EDE and DESI data (where the cosmological constant
remains consistent at $\lesssim 2\sigma$\cite{Wang:2024dka}), we
find that the EDE–DM coupling slightly favors evolving dark
energy in the axion-like EDE model, whereas the suppression
persists in the AdS-EDE scenario. Moreover, in the presence of
evolving dark energy, AdS-EDE exhibits a mild preference for $c >
0$ (with a larger mean value), in contrast to the axion-like EDE
case and to results obtained with pre-DESI BAO data. Notably,
AdS-EDE favors a non-zero coupling $f_* > 0$ at the $\gtrsim
1\sigma$ level.

This discrepancy suggests that the specific construction of the
EDE potential, beyond the mere existence of an EDE component, can
alter our understanding of the late-time universe when the
coupling between EDE and DM is taken into account. Our results
highlight the potential significance of a runaway EDE potential;
see Refs. \cite{Bernardo:2020lar,Bernardo:2021vfw,Wang:2025dtk}
for further model constructions.


\section*{Acknowledgments}

This work is supported by NSFC, No.12475064, National Key Research
and Development Program of China, No. 2021YFC2203004, and the
Fundamental Research Funds for the Central Universities. We
acknowledge the use of publicly available codes AxiCLASS
(\url{https://github.com/PoulinV/AxiCLASS}) and classmultiscf
(\url{https://github.com/genye00/class_multiscf.git}).

\end{document}